\documentclass[proof]{pasj00}
\Received{2013 July 18}
\Accepted{2013 September 13}
\SetRunningHead{M. Fukagawa et al.}{High Surface Density in the Ring Disk around HD~142527}

\usepackage{times}

\newcommand{\sun}{\ensuremath{\odot}}%
\newcommand{\icarus}{{\it Icarus}}%

\begin{document}

\title{Local Enhancement of Surface Density in the Protoplanetary Ring Surrounding HD~142527}
\author{Misato FUKAGAWA, \altaffilmark{1}
	Takashi TSUKAGOSHI, \altaffilmark{2}
	Munetake MOMOSE, \altaffilmark{2}
	Kazuya SAIGO, \altaffilmark{3}
	Nagayoshi OHASHI, \altaffilmark{4}
	Yoshimi KITAMURA, \altaffilmark{5}
	Shu-ichiro INUTSUKA, \altaffilmark{6}
	Takayuki MUTO, \altaffilmark{7}
	Hideko NOMURA, \altaffilmark{8}
	Taku TAKEUCHI, \altaffilmark{9}
	Hiroshi KOBAYASHI, \altaffilmark{6}
	Tomoyuki HANAWA, \altaffilmark{10}
	Eiji AKIYAMA, \altaffilmark{3}
	Mitsuhiko HONDA, \altaffilmark{11}
	Hideaki FUJIWARA, \altaffilmark{4}
    Akimasa KATAOKA, \altaffilmark{3,12}
    Sanemichi Z. TAKAHASHI, \altaffilmark{6,8}
	Hiroshi SHIBAI \altaffilmark{1}
}
\altaffiltext{1}{Graduate School of Science, Osaka University, 1-1 Machikaneyama, Toyonaka, Osaka 560-0043, Japan}\email{misato@iral.ess.sci.osaka-u.ac.jp}
\altaffiltext{2}{College of Science, Ibaraki University, 2-1-1 Bunkyo, Mito, Ibaraki 310-8512 Japan}
\altaffiltext{3}{National Astronomical Observatory of Japan, 2-21-1, Osawa, Miaka, Tokyo 181-8588, Japan}
\altaffiltext{4}{Subaru Telescope, 650 North A'ohoku Place, Hilo, HI, 96720, USA}
\altaffiltext{5}{Institute of Space and Astronautical Science, Japan Aerospace Exploration Agency, 3-1-1 Yoshinodai, Chuo, Sagamihara 252-5210, Japan}
\altaffiltext{6}{Department of Physics, Graduate School of Science, Nagoya University, Furo-cho, Chikusa-ku, Nagoya, 464-8601, Japan}
\altaffiltext{7}{Division of Liberal Arts, Kogakuin University, 1-24-2, Nishi-Shinjuku, Shinjuku-ku, Tokyo 163-8677, Japan}
\altaffiltext{8}{Department of Astronomy, Graduate School of Science, Kyoto University, Kyoto 606-8502, Japan}
\altaffiltext{9}{Department of Earth and Planetary Sciences, Tokyo Institute of Technology, Meguro-ku, Tokyo 152-8551, Japan}
\altaffiltext{10}{Center for Frontier Science, Chiba University, Inage-ku, Chiba 263-8522, Japan}
\altaffiltext{11}{Department of Mathematics and Physics, Kanagawa University, 2946 Tsuchiya, Hiratsuka, Kanagawa 259-1293, Japan}
\altaffiltext{12}{School of Physical  Sciences, Graduate University for Advanced Studies (SOKENDAI), Mitaka,  181-8588, Tokyo, Japan}


\maketitle

\begin{abstract}
We report ALMA observations of dust continuum, $^{13}$CO $J=$ 3--2, and C$^{18}$O $J=$ 3--2 line emission toward a gapped protoplanetary disk around HD~142527. The outer horseshoe-shaped disk shows the strong azimuthal asymmetry in dust continuum with the contrast of about 30 at 336~GHz between the northern peak and the southwestern minimum. In addition, the maximum brightness temperature of 24~K at its northern area is exceptionally high at 160~AU from a star. To evaluate the surface density in this region, the grain temperature needs to be constrained and was estimated from the optically thick $^{13}$CO $J =$ 3--2 emission. The lower limit of the peak surface density was then calculated to be 28 g cm$^{-2}$ by assuming a canonical gas-to-dust mass ratio of 100. This finding implies that the region is locally too massive to withstand self-gravity since Toomre's $Q \lesssim$1--2, and thus, it may collapse into a gaseous protoplanet. Another possibility is that the gas mass is low enough to be gravitationally stable and only dust grains are accumulated. In this case, lower gas-to-dust ratio by at least 1 order of magnitude is required, implying possible formation of a rocky planetary core. 
\end{abstract}

\section{Introduction}
To understand how planets form, it is critical and most straightforward to observe the actual process of planet building at their birthplaces \citep{bec90,wil11}. 
The key approach is to determine the detailed structure of a protoplanetary disk, which can provide an indicator of planet-forming activity \citep{mut12}. Great attention has therefore been paid to transitional disks with voids of material in the shape of holes or 
gaps \citep{str89,and11}. 
Such structures are often interpreted as a consequence of the occurrence of giant planets which can clear the disk along their orbits. From the other perspective, the outer, ring-like disk itself may serve as a cause of planet formation \citep{mat12,may12,pin12}. 

HD~142527 is a Herbig Fe star \citep{wae96} surrounded by a disk exhibiting a wide gap with a radial width of approximately 100~AU (\cite{fuk06, fuj06, ver11}, hereafter V11; \cite{ram12,cas12}). The distance is assumed to be 140~pc in this paper considering the association to Sco OB2. The stellar mass and the age are about 2~$M_{\sun}$ and  5 Myr, respectively,  estimated adopting 140--145~pc (\cite{fuk06}, V11). 
Most recently, faint stream-like features were found in 
HCO$^+$ and dust continuum at 345 GHz with ALMA, and these were interpreted as funnel flows into the inner disk through giant planets \citep{cas13}. Here, we revisit the disk structure based on our ALMA data with better sensitivity and angular resolution. We confirmed the presence of the inner disk, and in particular, emphasize the surprisingly high surface density of dust at 160 AU from the star in this letter.

\section{Observations and Data Reduction}
HD~142527 was observed with ALMA in Band 7 using 20--26 12-m antennas in the Extended array configuration in Cycle~0. The maximum and minimum baselines were 380~m and 20~m, respectively, and the latter corresponded to the largest  angular scale of the detectable component of $10\arcsec$. 
The observations reported in this letter consisted of four scheduling blocks  over the period from June to August 2012. 
The correlator was configured to store dual polarizations in four separate spectral windows with 469~MHz of bandwidth and 3840 channels each, providing a  channel spacing of 0.122~MHz (0.11~km~s$^{-1}$). Note that the effective spectral resolution is lower by a factor of around 2 ($\sim$0.2 km~s$^{-1}$) because of Hanning smoothing. The central frequencies for these windows are 330.588, 329.331, 342.883, and 342.400~GHz, respectively, allowing observations of molecular lines of $^{13}$CO $J=3$--2, C$^{18}$O $J=3$--2, and CS $J=7$--6. The results of the CS observations will be reported elsewhere. 
The quasars 3C~279 and J1924-292 were targeted as bandpass calibrators, whereas  the amplitude and phase were monitored through observations of the quasar J1427-4206.
The absolute flux density was determined from observations of Titan and Neptune. 

The data were calibrated and analyzed using the Common Astronomy Software Applications (CASA) package, version 3.4. After flagging aberrant data and applying calibrations for the bandpass, gain, and flux scaling, the corrected visibilities were imaged and deconvolved using the CLEAN algorithm with Briggs weighting with a robust parameter of 0.5. 
In addition, to improve the sensitivity and image fidelity, the self-calibration was performed for the continuum for which the distinct structure was detected with a very high signal-to-noise ratio (S/N). The calibration began with the CLEAN-ed image as an initial model of the source brightness distribution. The phase alone was corrected first via six iterative model refinements, then the calibration was obtained for phase-plus-amplitude with no iteration. 
The solution for the continuum was applied to $^{13}$CO and C$^{18}$O data. The final CLEANing was performed with Uniform weighting both for the continuum and emission lines. 
The self-calibration reduced the fluctuation in the continuum to the level that can be accounted for by 2--3 times as high as the theoretical thermal origin, resulting in a clear detection of compact emission at the stellar position. 

The uncertainty associated with the absolute flux density is 10\%. 
The synthesized beam size for the continuum is $0\farcs39 \times 0\farcs34$ with a position angle (PA) of 57$\arcdeg$ for the major axis, and those for $^{13}$CO and C$^{18}$O are $0\farcs43 \times 0\farcs37$ with PA = 50$\arcdeg$.
The rms noise is 0.19 mJy beam$^{-1}$  for the continuum whereas it is 12 and 15 mJy beam$^{-1}$ in the 0.11 km~s$^{-1}$ wide channels for the line emission of $^{13}$CO and C$^{18}$O, respectively. Since the positional information was lost through the self-calibration, the stellar position was determined as the brightness centroid of the compact continuum detected at around the stellar coordinates.

\section{Results}
\subsection{Continuum at 336 GHz}\label{sec:cont}
\subsubsection{Outer Disk}\label{sec:cont_outerdisk}
\begin{figure}
\begin{center}
\FigureFile(100mm,68mm){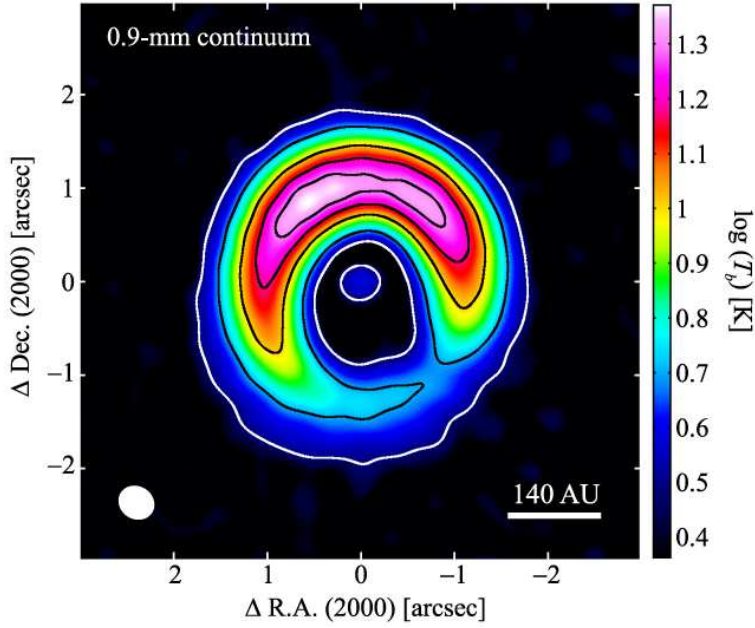}
\caption{Dust continuum map at 336 GHz (890~$\mu$m) for the disk of HD~142527. The color-scale shows the brightness temperature in a logarithmic scale. The $black$ contours denote $T_b$ = 5, 10, 15 and 20 K while the $white$ one does the 5$\sigma$ level. The size of the synthesized beam is shown at the left corner of the image with a white ellipse of $0\farcs39 \times 0\farcs34$ (= 55 AU $\times$ 48 AU) with the major-axis PA of $57\arcdeg$. 
\label{fig:dust}}
\end{center}
\end{figure}

Figure~\ref{fig:dust} shows the continuum emission at 336 GHz (890~$\mu$m). The outer disk was readily detected and the total flux density ($>5\sigma$) was measured as 2.7~Jy.
It clearly departs from a uniform ring and exhibits a horseshoe-like distribution as reported in previous studies \citep{oha08,cas13}. The northern region is brighter than the southwestern (SW) area, with an emission peak of 213~mJy~beam$^{-1}$ at the projected distance of $1\farcs0$ (140~AU) from the star, and at a PA of $30\arcdeg$. When probed along the annular emission ridge, both ends of the horseshoe appear to be connected at the brightness minimum of 7~mJy~beam$^{-1}$ at $1\farcs3$ (180~AU) and PA = 220$\arcdeg$. The contrast in flux density thus reaches to 30 between the northern peak and the SW minimum. 
The peak flux density is expressed by a brightness temperature ($T_b$) of 24 K  without the Rayleigh-Jeans approximation. This $T_b$ is much higher than those beyond 100~AU in other disks. 
Unless the emitting grains are significantly warmer than 24 K, the region should be optically thick to the submillimeter radiation due to the high column density. 

\subsubsection{Inner Disk and Radial Gap}\label{sec:innerdisk}
At the position of the star, emission was detected with a significance of 16$\sigma$ at the peak (figure~\ref{fig:dust}). 
The $\chi^2$-fitting of an  elliptic Gaussian function  resulted in a FWHM  of $(0\farcs33 \pm 0\farcs02) \times (0\farcs29 \pm 0\farcs02)$  with a PA of $99\arcdeg \pm 3\arcdeg$ for the major axis. It was thus spatially unresolved given the beam size of $0\farcs39 \times 0\farcs34$. 
The integrated flux density over the  Gaussian was 2.3$\pm$0.2~mJy,  substantially higher than the photospheric level of 0.02~mJy. This suggests that the emission comes from the inner disk whose presence has been predicted from the near-infrared excess, and the mid-infrared imaging (V11; \cite{fuj06}) where the inner disk was marginally resolved with an inferred size of $\sim$30--50~AU in radius. The gaseous emission from the inner disk was also imaged and kinematically resolved \citep{cas13,obe11,pon11}.
The mass of the inner disk can be crudely estimated, assuming that the majority of the mass resides in the outer, optically thin part,  using the equation $M_{\rm in} = F_{\nu} d^2 / (\kappa_{\nu} B_{\nu}(T))$, where $F_{\nu}$ is the observed flux density, $d$ is the distance to the star, $\kappa_{\nu}$ is the  opacity, and $T$ is the characteristic temperature of the disk. We adopted $T=50$~K \citep{chi97} and $\kappa_{336}$ of 0.034~cm$^2$ g$^{-1}$, assuming a gas-to-dust mass ratio (hereafter, we denote ``g/d") of 100, on the basis of  the conventional relation $\kappa_{\nu}=0.1(\nu / 10^{12}$~Hz)$^{\beta}$  with $\beta = 1.0$ (\cite{bec90}; V11); the total (gas and dust) disk mass was then calculated as $(4.3 \pm 0.4) \times 10^{-5}~M_{\sun}$. Note that a considerable uncertainty exists in the assumption of optical thickness, and the mass derived here can give the lower limit. 
In the previous spectral energy distribution modeling, the dust (not including gas) mass of the inner disk was estimated to be $2.5 \times 10^{-9} M_{\sun}$, but the model was constrained based primarily on the near- and mid-infrared excess and assumed small, micron-sized grains (V11).
The flux density at 890~$\mu$m obtained with ALMA seems about one order of magnitude higher than that expected in their modeling.
The detection of the submillimeter continuum in our imaging suggests that the bulk of the mass resides in larger grains.

In the radial gap, the surface brightness decreases to the background level.
Dust streamers from the outer disk reported in the previous study \citep{cas13} were not confirmed.

\begin{figure}
\begin{center}
\FigureFile(100mm,68mm){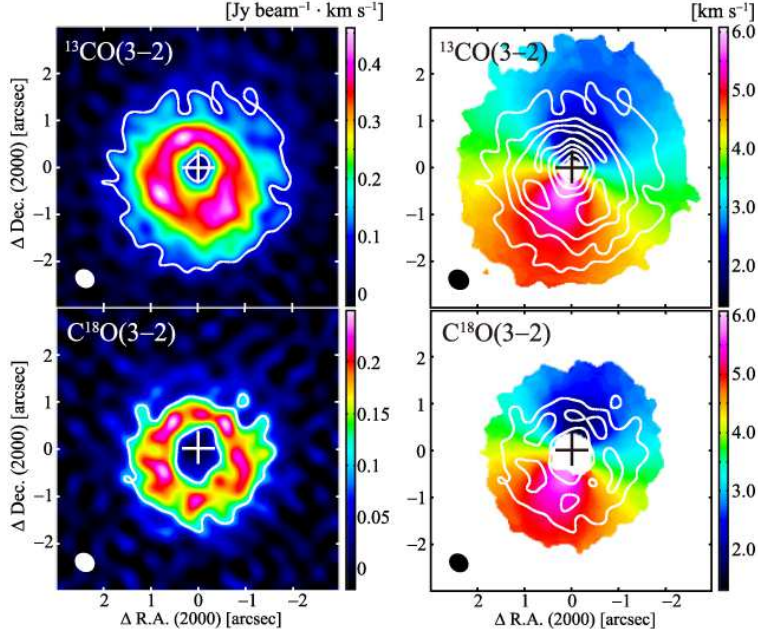}
\caption{Moment 0 ($left$) and 1 ($right$) maps in $^{13}$CO ($upper$) and C$^{18}$O $J=$~3--2 ($lower$). The contours in the $left$ panels show the 5$\sigma$ levels. In the $right$ panels, the first moment maps are presented for the emission detected above 5$\sigma$. The contours for the integrated intensity are overplotted, starting from 5$\sigma$ and increasing by 5$\sigma$ steps. The $cross$ in each panel denotes the position of the central star. 
The synthesized beam size is displayed with a white ellipse in each panel, and is  $0\farcs43 \times 0\farcs37$ with a PA of $50\arcdeg$ both for $^{13}$CO and C$^{18}$O. \label{fig:co}} 
\end{center}
\end{figure}

\subsection{$^{13}$CO $J=$~3--2 and C$^{18}$O $J=$~3--2}
\subsubsection{Integrated Intensity and Gas Kinematics} \label{sec:gas_totalintensity}
The left panels of figure~\ref{fig:co} present integrated intensity (0th-moment) maps of the $^{13}$CO $J=$~3--2 and C$^{18}$O $J=$~3--2 line emission. 
The brightness distribution does not show as strong an asymmetry in the azimuthal direction as that observed in the dust continuum, and fluctuate by a factor of less than 2. In the radial direction, the integrated intensity has peaks at $r = 0\farcs7$--$1\farcs1$ depending on the azimuthal angle in $^{13}$CO, and at $r = 0\farcs8$--$1\farcs2$ in C$^{18}$O. No emission was detected above 3$\sigma$ around the stellar position in both lines (1$\sigma = 18.9$ and 19.6~mJy beam$^{-1}$$\cdot$km s$^{-1}$ for $^{13}$CO and C$^{18}$O, respectively). 

The right panels of figure~\ref{fig:co} show 1st-moment maps in $^{13}$CO and C$^{18}$O. 
Despite the highly structured continuum, the velocity fields are consistent with a simple, circular Keplerian rotation within the resolution of 0.2~km s$^{-1}$, which was confirmed as follows. The position-velocity relation was extracted along the major axis (PA = $-19\arcdeg \pm 2\arcdeg$), and the peak velocity was estimated by the Gaussian fitting at each radius every $0\farcs03$. Then, S/N-weighted least-squares fitting of an analytic Keplerian equation was performed to the measured peak velocity as a function of radius. 
In the fitting, the systemic velocity, position of the center of mass (the star), and inclination relative to an observer are set as free parameters, whereas the stellar mass is fixed in the range of $2.2 \pm 0.3$~$M_{\sun}$ (V11). 
Using the $^{13}$CO data with a higher S/N, the inclination was estimated as $i = 26.9^{+2.2}_{-1.8}$ degrees, where the uncertainty is dominated by the error in the adopted stellar mass. Note that $i$ is not large enough to yield reasonable ($\Delta i < \pm10\arcdeg$, $\Delta M_* < \pm50$\%)  constraints on both the stellar mass and the inclination \citep{sim00}. The  systemic velocity was determined as $3.70 \pm 0.02$~km~s$^{-1}$. The location of the velocity centroid matches with the compact component of the continuum emission within $0\farcs04$. 

\subsubsection{Temperature Estimate}
\begin{figure}
\begin{center}
\FigureFile(100mm,45mm){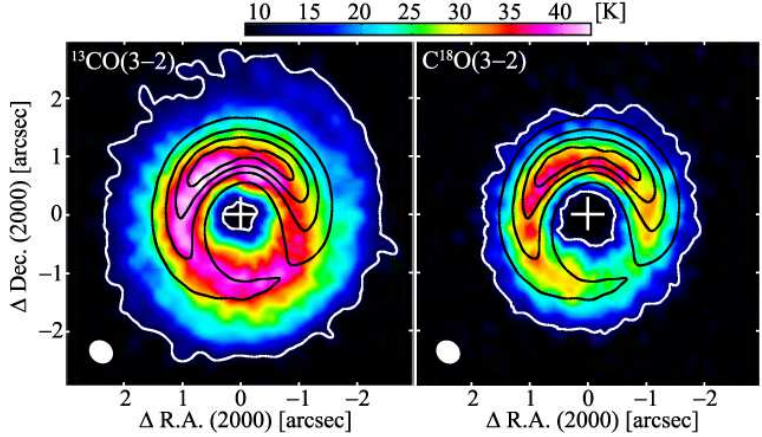}
\end{center}
\caption{Peak intensity maps of $^{13}$CO ($left$) and C$^{18}$O $J=$~3--2 ($right$). The peak intensity denoted by color includes the underlying continuum. The $white$ contour shows the 5$\sigma$ level. The $T_b$ for the continuum is also plotted with $black$ contours indicating 5, 10, 15, and 20~K. The $cross$ denotes the stellar position. The faint arm-like feature is seen in the $^{13}$CO map in northwest at the disk outer edge, which corresponds to the arm detected in scattered light \citep{fuk06}. 
\label{fig:peakintensity}}
\end{figure}

The line results above were obtained after subtracting the underlying continuum, which is the typical method adopted in earlier studies. However, the continuum from the outer disk of HD~142527 shows the strong 
asymmetry with the high $T_b$ (Section~\ref{sec:cont_outerdisk}). 
In fact, the peak intensity maps of $^{13}$CO and C$^{18}$O most evidently show the flux deficits in north.
When the $T_b$ map of the continuum is added to that of the line peak intensity, after matching the beam size to that of the line data, the resultant distribution of $T_b$($^{13}$CO) is ring-like and azimuthally uniform (figure~\ref{fig:peakintensity}). In radial direction, the maximum $T_b$($^{13}$CO) is 41~K at 40~AU inside the peak of the continuum, when measured in the deprojected ($i=27\arcdeg$) profile averaged over the PA range for the bright continuum, from $-49\arcdeg$ to $51\arcdeg$. The $T_b$($^{13}$CO) at the location of the continuum peak is 36~K, which happens to coincide with the highest $T_b$(C$^{18}$O) displaced inward by 28~AU from the continuum peak.

The optical depth of $^{13}$CO is greater than unity over the entire disk detected above 5$\sigma$ in C$^{18}$O, judging from the continuum-subtracted ratio between the C$^{18}$O peak intensity and the $^{13}$CO intensity at the same velocity as C$^{18}$O. Here, we assume the same excitation temperature ($T_{\rm ex}$) for both lines and an abundance ratio of X(C$^{13}$O)/X(C$^{18}$O) $\sim$ 7 \citep{qi11}. 
In addition, except for the SW region, $T_b$(C$^{18}$O) is systematically lower by $\sim$5~K than $T_b$($^{13}$CO) in the outer region beyond the radial $T_b$ peak, and 
this relation can be naturally understood by the situation that the line emission comes from optically thick surfaces in the upper and lower layers  for $^{13}$CO and C$^{18}$O, respectively. Moreover, the radial positional shifts of the highest $T_b$ for the lines from that for the continuum can be attributed to the inner, warmer emitting surfaces for the gaseous component exposed to the central star.
Therefore, the line intensity, at least for $^{13}$CO, reflects the physical (kinetic) temperature, not the column density under local thermodynamic equilibrium (LTE). The emission can be approximated by the LTE conditions because the density discussed here (see Section~\ref{sec:sd}) is well above the critical densities for $^{13}$CO and C$^{18}$O $J=$~3--2 \citep{pav07}. When estimating a temperature of the optically thick surface, the continuum needs to be added since it is non-negligible for this object.

\section{Discussion}
\subsection{Constraint on Surface Density}\label{sec:sd}
The bulk of grains are expected closer to the disk mid-plane, and it is unlikely that they are warmer than the emitting surfaces in $^{13}$CO and C$^{18}$O $J=$~3--2. 
Therefore, we conservatively regard 36~K, which is the $T_b$($^{13}$CO) at the radius of continuum peak, as an upper bound of the grain temperature ($T_{\rm dust}$). A lower limit of the surface density can thus be estimated with $T_{\rm dust}$=36~K. 
The dust optical depth ($\tau$) was derived by comparing $T_{\rm dust}$ with the continuum $T_b$. In the northern horseshoe peak, using $\tau = 0.9$ along with $\kappa_{336} = 0.034$~cm$^2$ g$^{-1}$ for a g/d of 100 (Section~\ref{sec:innerdisk}) , the surface density was calculated as $\Sigma_{\rm peak}=$ 28.1~g~cm$^{-2}$. In the SW minimum, the same assumption of 36~K resulted in $\Sigma_{\rm min} =  0.7$~g~cm$^{-2}$, giving a density contrast of 40 relative to the northern peak. 
Integrating the surface density 
yielded a mass for the entire outer disk of 0.09~$M_{\odot}$, which turns out to be consistent with the previous estimate of 0.1~$M_{\sun}$ obtained by the conventional, optically thin prescription with 40~K (V11).  

Because $T_b \approx T_{\rm dust} \cdot \tau$, 
  grains must be warmed in the dense region near the mid-plane to account for the observed high $T_b$. We checked if this condition can be realized  
by employing radiative transfer calculations along the disk vertical direction. The 1+1D modeling \citep{nom09} was performed assuming the surface density estimated as above, 
vertical hydrostatic equilibrium, and the incident stellar luminosity of $15~L_{\sun}$ considering the possible shadowing by the inner disk (V11). The  observed $T_b$ was reproduced when using a grain opacity plausible for protoplanetary disks \citep{nom05}.

The optical depth of C$^{18}$O may fall below unity in the SW region given the lower $T_b$ (figure~3). We thus attempted to estimate the gas surface density using a one-zone approximation in the vertical direction. 
The optical depth was derived from the line ratio of C$^{18}$O to $^{13}$CO with an abundance ratio taken from \citet{qi11}, and the temperature was provided by the $T_b$($^{13}$CO) 
 (figure~3).
The calculation inferred the g/d of 14 at PA = $210\arcdeg$--$230\arcdeg$. 
However, the optical depth was suggested to be about 1.0. In addition, we cannot completely exclude the possibility that the C$^{18}$O emission is more optically thick with a lower $T_{\rm ex}$. 
Thus, the derived g/d can provide a lower limit.

\subsection{Spatial Structure}\label{sec:structure}
\begin{figure}
\begin{center}
\FigureFile(100mm,60mm){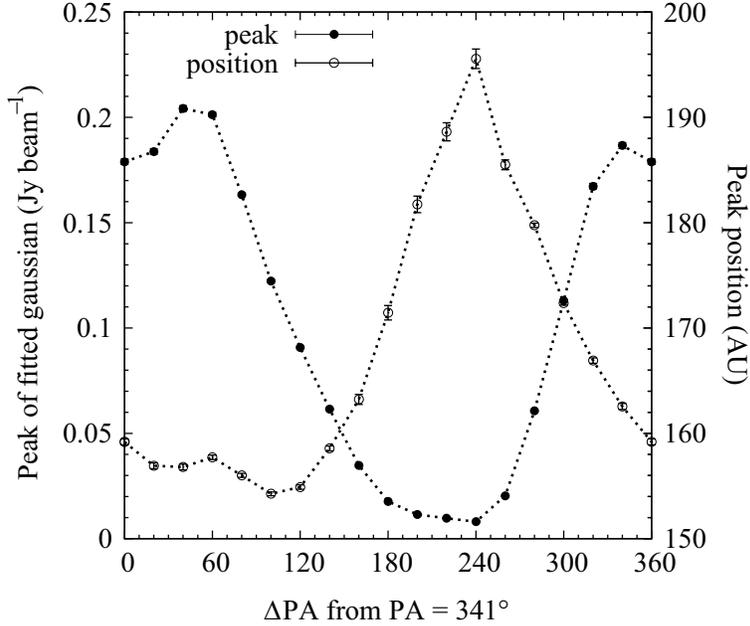}
\end{center}
\caption{Peak intensity and its location from the star obtained by the Gaussian fitting to the profile averaged over every  20$\arcdeg$ bin in PA. The PA in this plot is measured from the major axis (PA=$-19\arcdeg$). \label{fig:rprof}} 
\end{figure}

The radial profile for the continuum emission is well described by a Gaussian function, rather than a power law, at all azimuthal angles. 
A Gaussian fitting  to the deprojected, azimuthally-averaged profile gave $I_{\rm{average}} \rm{(Jy~beam}^{-1}\rm{)}= 9.95 \times 10^{-2} \exp\left[-\{(R-161.1)/(48.2)\}^2\right]$, where $R$ is the distance from the star in AU. 
Assuming a g/d of 100 and a temperature of 36~K, the fitted Gaussian to the azimuthally-averaged one has a peak surface density, $\Sigma_{\rm average}$, of 11.3~g~cm$^{-2}$ at $R = 161$~AU. 
The Gaussian fitting to the radial profile in each PA yielded one clear feature, which is the anti-correlation  between the peak intensity of the Gaussian and its deprojected distance from the star (figure~\ref{fig:rprof}). The anti-correlation excludes the possible explanation that the azimuthal asymmetry is due to eccentric orbits of the dust particles. If the orbits are eccentric, they concentrate near the apastron. 

If the $density$ profile is confirmed to be Gaussian after incorporating a realistic temperature gradient, it is natural to understand the disk as a ring or torus. In a primordial disk, the radial surface density distribution can be expressed as a power law accompanied by an outer, exponential tapered edge \citep{hug08}. An inner gap or hole can then be created by being carved by a planet(s) for instance, but the outer boundary should remain unaffected unless the gap stretches to near the disk outer edge or the source experiences stellar encounters. 
It would be worth noting that formation of a ring can be explained by different mechanisms without the aid of planets, such as a secular process through viscous overstability (Takahashi \& Inutsuka in prep., \cite{sch95}). 

On the other hand, the $^{13}$CO emission was detected out to $r$$\sim$350~AU on average at all PAs (5$\sigma$, figure~3). This large extent is not reconciled with the Gaussian profile for the continuum. The observed $^{13}$CO intensity at 350~AU is several orders of magnitude higher than that predicted by an extrapolation from the Gaussian for any excitation temperature assuming LTE and a g/d of 100. 
This suggests that there is an additional floor spreading outward. The non-detection of the dust continuum for this floor is not inconsistent with the assumption of a g/d of 100 considering the detection limit in our observations. 

\subsection{Stability against Self-Gravity}
What can be expected from the local density enhancement in a Keplerian disk? 
To evaluate the gravitational stability, the Toomre's $Q$ parameter \citep{too64} was computed under the assumption of a g/d of 100. 
It was first estimated toward the averaged surface density to examine the global stability, resulting in $Q(R = 161~{\rm{AU}}) = 2.2$ using $\Sigma_{\rm{average}} = 11$~g cm$^{-2}$, the isothermal sound speed for 36~K, and a stellar mass of $2.2~M_{\sun}$. Locally estimated at the horseshoe peak at $R = 156$~AU, $Q$ was obtained as 0.9 for $\Sigma_{\rm peak} = 28$~g cm$^{-2}$. Note that $\Sigma$ is at its lower limit, whereas the temperature (sound speed) is the upper bound; therefore, the obtained $Q$ is considered to provide the upper limit. $Q \lesssim$ 1--2 indicates that the disk is vulnerable to gravitational instability if a g/d is $\sim$100. 
Crudely assuming that a resultant fragment acquires the mass within a local volume determined by the size of the disk scale height (16~AU at 36~K), it gains $\sim$3 Jupiter-masses.
The dynamical clumping of the ring or torus structure may be considered in such a way that filaments fragment into stars in molecular clouds (\cite{inu92}, 1997). 
Note that the disk fragmentation likely occurs on a dynamical timescale ($\sim$$10^3$ years), which is much shorter than the stellar age ($\sim$$10^6$ years), suggesting that we are observing HD 142527 in its transient phase. 

The caveat in the above discussion is the assumption of the g/d (e.g., \cite{til12}). 
For instance, if a g/d is at the lower end of that estimated in the SW region ($\sim$10, Section~\ref{sec:sd}) or even smaller, the Toomre's Q becomes $\gg$1 and the disk can be stable against its self-gravity. This situation is allowed since C$^{18}$O emission is expected to stay optically thick at the horseshoe peak even if a g/d is as small as the order of unity. 
Such a g/d, lower by more than a factor of $\sim$10 than the ISM value, suggests grain accumulation. As a result, the process may lead to the efficient growth of a solid planetary core.
The phenomenon is consistent with the scenario of particle trapping at pressure maxima that can be caused by several possible mechanisms, including a perturbation by an already formed planet \citep{reg12,bir13}.

In either case, the  pile-up of disk material beyond 100 AU 
is quite surprising in the classical scenario of planet formation. HD~142527 is indeed unique compared to other ring-like disks that have been spatially resolved. They show relatively smaller disk masses ($\sim$1\% of $M_*$ for a g/d of 100), and their azimuthal brightness fluctuations in the continuum are lower by an order of magnitude \citep{and11,mat12,ise10,tan12}. In terms of strong asymmetry, the similarity can be found only in the recent discovery of the crescent-shaped disk 
with the azimuthal contrast in flux density of at least 130 at 680~GHz \citep{mar13}. 
Therefore, HD~142527 offers a rare opportunity for us to directly observe the critical moment of planet formation and can provide new insights into the origin of wide-orbit planetary bodies \citep{mar08,ire11}

\bigskip

We are grateful to the referee for the valuable comments which improved our manuscript. We also thank Y. Aikawa for the helpful discussion. 
This paper makes use of the following ALMA data: ADS/JAO.ALMA\#2011.0.00318.S. ALMA is a partnership of ESO (representing its member states), NSF (USA) and NINS (Japan), together with NRC (Canada) and NSC and ASIAA (Taiwan), in cooperation with the Republic of Chile. The Joint ALMA Observatory is operated by ESO, AUI/NRAO and NAOJ. This work is supported by MEXT KAKENHI No. 23103004, 23103005.

\end{document}